\documentstyle[prb,eqsecnum,aps,epsf]{revtex} 

\tighten

\begin{document}
\draft \title{Quantum interference and Coulomb interaction in arrays
of tunnel junctions.}  \author{I.S. Beloborodov$^{(1)}$,
K.B. Efetov$^{(1,2)}$, Alexander Altland$^{(1)}$, and F.W.J.
Hekking$^{(3)}$} \address{$^{(1)}$Theoretische Physik III,
Ruhr-Universit\"{a}t Bochum, 44780 Bochum, Germany\\
$^{(2)}$L.D. Landau Institute for Theoretical Physics, 117940 Moscow,
Russia\\ $^{(3)}$ Laboratoire de Physique et Mod\'elisation des
Milieux Condens\'es, Maison des Magist\`eres Jean Perrin\\ BP166,
38042 Grenoble Cedex 9, France} \date{\today} \maketitle

\begin{abstract}
  We study the electronic properties of an array of small metallic
  grains connected by tunnel junctions. Such an array serves as a
  model for a granular metal. Previous theoretical studies of junction
  arrays were based on models of quantum dissipation which did not
  take into account the diffusive motion of electrons within the
  grains.  We demonstrate that these models break down at sufficiently
  low temperatures: for a correct description of the screening
  properties of a granular metal at low energies the diffusive nature
  of the electronic motion within the grains is crucial.  We present
  both a diagrammatic and a functional integral approach to analyse
  the properties of junction arrays. In particular, a new effective
  action is obtained which enables us to describe the array at
  arbitrary temperature. In the low temperature limit, our theory
  yields the correct, dynamically screened Coulomb interaction of a
  normal metal, whereas at high temperatures the standard description
  in terms of quantum dissipation is recovered.

\end{abstract}


\section{Introduction}

During the past two decades much attention has been devoted to the
study of disordered bulk metals with electron-electron
interactions~\cite{Altshuler,Lee,Fin} from the one hand and arrays of
normal metallic grains connected by tunnel
junctions~\cite{Ben-Jacob,Zaikin,Fazio} from the other. The Coulomb
interaction plays a very important role in both types of system and
many non-trivial physical effects occur due to it. In disordered
metals, {\em e.g.}, the Coulomb interaction results in quantum
corrections to conductivity and reduces the density of states, a
phenomenon known as zero bias anomaly.  Qualitatively similar effects
were predicted~\cite{Fazio} and studied
experimentally~\cite{Mooij90,Tighe93,Yamada93,Kobayashi95} for arrays
of tunnel junctions in the normal state, where a sufficiently strong
Coulomb interaction can suppress the conductivity (Coulomb blockade of
tunneling).

In spite of the similarities in the physical behavior of disordered
metals and normal arrays of tunnel junctions, completely different
approaches are used for the theoretical description of these
respective systems.  As concerns disordered metals with interaction,
either diagrammatic methods~\cite{Altshuler} or non-linear $\sigma
$-models \cite{Fin82,Kamenev,Chamon} are employed.  The starting point
of these theories is a model of a weakly disordered bulk metal. On the
other hand, for the description of Coulomb blockade effects in a
junction array or a granular material, a completely different approach
based on the theory of Ambegaokar, Eckern and Sch\"{o}n
(AES)~\cite{Ambeg} is commonly used.  The free energy functional
proposed by AES does not contain any disorder: the finite conductance
obtained within this model is due to tunneling of electrons between
grains and can be incorporated into the theory through a term
describing the so-called quantum dissipation, first introduced
phenomenologically in Ref.~\onlinecite{Cal}.

Although both approaches have been introduced almost 20 years ago,
only a few indirect attempts have been made to reconcile them.  We
mention the work by Nazarov~\cite{Naz89} and Levitov and
Shytov~\cite{Lev97} that treat the diffusive zero bias anomaly and the
Coulomb blockade of tunneling on the same footing. More recently,
Nazarov~\cite{Naz99} showed how Coulomb blockade phenomena survive in
diffusive systems.  Indeed, the properties of a granular metal without
electron-electron interactions are qualitatively similar to those of a
disordered bulk metal (see e.g. Ref.~\onlinecite{Efetov}) and it would
be natural to expect that this qualitative similarity persists even in
the presence of interactions. In fact, one may conjecture that bulk
disordered metals from the one hand, and granular metals or arrays of
tunnel junctions from the other, can be described within one unifying
scheme.

In order to develop such a scheme, we consider in this paper a
granular normal metal with Coulomb interaction between electrons. The
consideration is simplified by taking into account only the long range
part of the interaction that leads to Coulomb blockade effects. We
assume that the macroscopic dimensionless conductance $g_{T}$ of the
granular metal is large, which enables us to develop a perturbative
theory with $g_{T}^{-1}$ as the small parameter. Within this theory,
physical quantities of interest can be calculated at arbitrary
temperatures. The diagrammatic analysis is supplemented by the
derivation of a $\sigma $-model that can be considerably simplified
provided one does not take into account weak localization effects.

We demonstrate below that granular metals and networks of tunnel
junctions can always be described using the standard techniques
developed in the theory of disordered metals (expansions in cooperons
and diffusons, $\sigma $- model calculations).  With the help of these
techniques one can in principle calculate any physical quantity
without using the notion of quantum dissipation.  Disorder is
inevitably present in the system even if electrons move ballistically
within the grains. In the latter case any small irregularity in the
shape of the grains causes the electron motion within the grains to be
chaotic and this assumption is sufficient for our theory to be
applicable.  An integrable shape of the grains does not seem to be
realistic, and even if it happened for isolated grains, tunneling from
grain to grain would add an additional chaoticity. These ideas about
intrinsic disorder in granular metals have been used in previous
studies~\cite{Efetov,Bel,Belobor}.

We find that, at not too low a temperature, the AES free energy
functional gives correct results.  However in the limit $T\rightarrow
0$ AES theory is no longer applicable.  The region of validity of the
AES free energy functional is determined by the inequality $T\gg
g_{T}\delta $, where $\delta =\left( \nu V\right) ^{-1}$ is the mean
level spacing, $\nu $ is the density of states at the Fermi surface
and $V$ is the volume of a single grain.

We conclude in particular that the AES notion of quantum dissipation
at $T=0$ can only be applied to tunnel junctions connecting infinitely
large conductors. If the volume of the conductors is finite the AES
picture can be applied at finite temperatures only. Apparently, the
notion of quantum dissipation should be treated with care. A
qualitative discussion of the problem ``dissipation versus dynamical
screening'' in the case of superconducting grains can be found in a
recent lecture by Altshuler\cite{Alt}.

The remainder of the paper is organized as follows. In
Sec.~\ref{choice} we formulate the model and write physical quantities
in terms of functional integrals over anticommuting fields. We
decouple the Coulomb interaction by an integration over auxiliary
fields. In Sec.~\ref{AES} we consider the AES action and show that it
does not correspond to the physics of normal metals with a screened
Coulomb interaction. In Sec.~\ref{metal} we develop a diagrammatic
technique. We show what kind of diagrams correspond to the AES action
and demonstrate that, at low temperatures, additional contributions
arise. Summing up these additional diagrams we arrive at expressions
corresponding to the dynamically screened Coulomb interaction of a
normal metal. Sec.~\ref{sigma} contains a derivation of a new action
for a granular metal with electron-electron interaction which is
applicable at arbitrary temperatures. The results are compared with
the results obtained diagrammatically and with those obtained on the
basis of the AES action. Our results are summarized in the Conclusion.

\section{Choice of the model}

\label{choice}

We consider an array of normal metallic grains coupled to each
other. The dimensionality of the array may be arbitrary.  The grains
are assumed to contain imperfections: there can be impurities inside
them as well as on their surface. This implies that the electron
motion in the grains is chaotic. The mean level spacing in a single
grain is equal to $\delta =\left( \nu V\right) ^{-1}$. However, the
shapes of the grains need not considerably differ from each other and
we assume for simplicity that the grains are arranged in a cubic
lattice.  The electrons can hop from grain to grain; moreover, they
interact with each other.  We are interested only in the long range
part of the Coulomb interaction and write it in a simplified form
describing charging of the grains. In such a formulation the spin of
the electrons is not important and can be taken into account at the
end of the calculations when writing proper densities.

Under these assumptions the electron motion can be conveniently described by
a functional integral of the type 
\[
\int \exp \left( -S[\psi ]\right) {\cal D}\psi ,
\]
with an action $S\left[ \psi \right] $ that can be written in the form 
\begin{equation}
S[\psi ]=S_{{\rm g}}[\psi ]+S_{{\rm t}}[\psi ]+S_{{\rm c}}[\psi ].
\label{a1}
\end{equation}
The action $S_{{\rm g}}\left[ \psi \right] $ in Eq.~(\ref{a1})
describes the electron motion within the grains and we write it in the
form
\begin{equation}
S_{{\rm g}}[\psi ]=\sum_{i}\int_{0}^{\beta }drd\tau \psi _{i}^{\ast }(\tau
,r)\left( \partial _{\tau }-\mu +\hat{H}_{i}\right) \psi _{i}(\tau ,r),
\label{a2}
\end{equation}
where $\psi _{i}\left( \tau ,r\right) $ is a fermion field in grain
$i$ at imaginary time $\tau $ and coordinate $r$ ($\psi ^{\ast }$ is
its complex conjugate), $\mu $ is the chemical potential, and $\beta
=T^{-1}$ is the inverse temperature.  The fermion fields $\psi
_{i}\left( \tau \right) $ satisfy the condition
\begin{equation}
\psi _{i}\left( \tau \right) =-\psi _{i}\left( \tau +\beta \right) .
\label{a51}
\end{equation}

The operator $\hat{H}_{i}$ in Eq. (\ref{a2}) includes scattering by
impurities and can be written as
\begin{equation}
\hat{H}_{i}=-{\frac{1}{2m}}{\bf \nabla }_{r}^{2}+u_{i}(r),  \label{a3}
\end{equation}
where $u_{i}\left( r\right) $ is the impurity potential. We assume that it
is Gaussian distributed with the correlation 
\begin{equation}
\langle u_{i}(r)u_{j}(r^{\prime })\rangle ={\frac{1}{2\pi \nu \tau
_{\rm imp}}} \delta (r-r^{\prime })\delta _{ij} . \label{a30}
\end{equation}

The action $S_{{\rm t}}\left[ \psi \right] $ stands for tunneling
between the grains, 
\begin{equation}
S_{{\rm t}}[\psi ]=\sum_{ij}\int_{0}^{\beta }d\tau \psi _{i}^{\ast }(\tau
)t_{ij}\psi _{j}(\tau ),  \label{a4}
\end{equation}
and $S_{{\rm c}}\left[ \psi \right] $ describes the charging of the grains 
\begin{equation}
S_{{\rm c}}[\psi ]=\frac{e^{2}}{2}\sum_{ij}\int_{0}^{\beta }d\tau
n_{i}(\tau )C_{ij}^{-1}n_{j}(\tau ), \label{a5}
\end{equation}
where 
\[
n_{i}\left( \tau \right) =\int dr\psi _{i}^{\ast }(\tau ,r)\psi _{i}(\tau
,r) 
\]
is the density field in grain $i$, $e$ is the electron charge, and
$C_{ij} $ is the capacitance matrix.  Eqs. (\ref{a1}~-~\ref{a5})
describe the model completely and one can start explicit calculations.

Due to the Coulomb interaction, the action $S[\psi ]$ is not quadratic
in $\psi $ and the integration over $\psi $ cannot be performed
immediately. A convenient way to proceed in such a case is to decouple
the term $S_{{\rm c}}[\psi ]$ by a Gaussian integration over auxiliary
fields. This transformation can be written as follows:
\begin{equation}
\int {\cal D}(\psi ^{\ast },\psi )e^{-S[\psi ]}={\cal N}\int {\cal D}
Ve^{-S_{2}[V]}\int {\cal D}(\psi ^{\ast },\psi )e^{-S_{1}[\psi ,V]},
\label{a6}
\end{equation}
where ${\cal N}$ is a normalization factor and $V_{i}\left( \tau
\right) $ is the decoupling field.  The action $S_{2}[V]$ in Eq.
(\ref{a6}) has the form
\begin{equation}
S_{2}[V]=\frac{1}{2}\int_{0}^{\beta }d\tau \sum_{ij}V_{i}(\tau
)C_{ij}V_{j}(\tau ),  \label{a7}
\end{equation}
which shows that the variable $V_{i}\left( \tau \right) $ has the meaning of
a voltage on grain $i$ at the time $\tau $. The new effective action 
$S_{1}[\psi ,V]$ reads 
\[
S[\psi ,V]=S_{{\rm g}}[\psi ,V]+S_{{\rm t}}[\psi ] , 
\]
with 
\begin{equation}
S_{{\rm g}}[\psi ,V]=\sum_{i}\int_{0}^{\beta }d\tau \psi _{i}^{\ast
}(\tau )\left( \partial _{\tau }-\mu +\hat{H}_{i}+ieV_{i}(\tau
)\right) \psi _{i}(\tau ) . \label{a8}
\end{equation}
One more transformation can be performed exactly. Following the procedure of
Ref.~\onlinecite{Gefen} where a single grain was considered, we represent 
$V_{i}\left( \tau \right) $ in the form 
\begin{equation}
V_{i}\left( \tau \right) =V_{i}^{0}+\widetilde{V}_{i}\left( \tau \right) ,
\label{a9}
\end{equation}
where $V_{i}^{0}$ is the static part and $\int_{0}^{\beta }
\widetilde{V}_{i}\left( \tau \right) d\tau =0$. The replacement 
\begin{equation}
\psi _{i}\left( \tau \right) \rightarrow \psi _{i}\left( \tau \right) \exp
\left( -i\phi _{i}\left( \tau \right) \right)  \label{a90}
\end{equation}
with 
\begin{equation}
\phi \left( \tau \right) =e\int_{0}^{\tau }\widetilde{V}_{i}\left( \tau
^{\prime }\right) d\tau ^{\prime }  \label{a10}
\end{equation}
does not violate the condition given by Eq. (\ref{a51}) and we can
remove the variable $\widetilde{V}_{i}\left( \tau \right) $ from the
action $S_{{\rm g}}[\psi ,V]$, Eq. (\ref{a8}). However, this variable
appears in the action $S_{{\rm t}}[\psi ]$ as an additional phase of
the field $\psi \left( \tau \right) $. As concerns the static part
$V_{i}^{0}$, its fluctuations can be neglected even in a single
isolated grain, provided the temperature is high $T\gg \delta
$~\cite{Gefen}.  This restriction is not necessary for the system of
the coupled grains in the limit of large conductance $g_{T}\gg 1$
considered here.

Using Eqs. (\ref{a1}- \ref{a10}) we finally reduce the calculation of
physical quantities to the computation of a functional integral of the
form
\begin{equation}
\int \exp \left( -S_{0}[\psi ]-S_{1}[\psi ,\phi ]-S_{2}[\phi ]\right) {\cal D
}\psi {\cal D}\phi ,  \label{a11}
\end{equation}
where 
\begin{equation}
S_{0}[\psi ]=\sum_{i}\int_{0}^{\beta }d\tau \psi _{i}^{\ast }(\tau )\left(
\partial _{\tau }-\mu +\hat{H}_{i}\right) \psi _{i}(\tau ),  \label{a12}
\end{equation}
\begin{equation}
S_{1}[\psi ,\phi ]=\sum_{ij}\int_{0}^{\beta }d\tau t_{ij}\psi _{i}^{\ast
}(\tau )\psi _{j}(\tau )\exp \left( i\phi _{ij}\left( \tau \right) \right) ,
\label{a13}
\end{equation}
\begin{equation}
S_{2}[\phi ]=\int_{0}^{\beta }d\tau \sum_{ij}\frac{C_{ij}}{2e^{2}}\frac{
d\phi _{i}(\tau )}{d\tau }\frac{d\phi _{j}(\tau )}{d\tau },  \label{a14}
\end{equation}
and $\phi _{ij}\left( \tau \right) =\phi _{i}\left( \tau \right) -\phi
_{j}\left( \tau \right) $.

Eqs. (\ref{a11}-\ref{a14}) completely specify the model that will be
studied
in the subsequent sections. Disorder is still present in the operators $\hat{
  H}_{i}$, Eq. (\ref{a3}), but the mean free path within the grains is
assumed to be large, $lk_{0}\gg 1$, where $k_{0}$ is the Fermi
momentum. In this limit, the macroscopic conductivity is determined
mainly by the tunneling conductance $g_{T}$,
\begin{equation}
g_{T}=\pi /(2e^{2}R_{T})=2\pi ^{2}t^{2}\nu ^{2}\approx 6.45k\Omega /R_{T}
\label{Sg}
\end{equation}
where $R_{T}$ is the tunneling resistance.

\section{ Quantum dissipation description of granular metals with
Coulomb interaction}

\label{AES}

A functional integral formulation was used in Ref.~\onlinecite{Ambeg} 
to treat the quantum dynamics of a Josephson junction. 
The action derived in that work (AES action) was used later 
in the context of a normal tunnel junction~\cite{Ben-Jacob,Zaikin}, as well as
in connection with arrays of tunnel junctions
in a number of  papers~\cite{Fazio}. 
The AES action can be obtained from 
Eqs.~(\ref{a11}- \ref{a14}). First, one should
average the Green function $G_{0}$ corresponding to the action 
$S_{0}[\psi ]$, Eq. (\ref{a12}), over impurities in the first 
Born approximation,
which gives (in Fourier representation) 
\begin{equation}
G_{0\varepsilon }\left( {\bf p}\right) =\left( i\varepsilon -\xi({\bf p})+
i\frac{\mbox{sgn}\left( \varepsilon \right) }{2\tau _{\rm imp}}\right) ^{-1} , \label{a15}
\end{equation}
where $\tau _{\rm imp}$ is defined in Eq. (\ref{a30}) and $\xi({\bf
  p})={\bf p}^2/2m -\mu$.  After that one should make a cumulant
expansion in the term $S_{1}[\psi,\phi ]$, Eq. (\ref{a13}). Keeping
only the second order of the expansion and assuming for simplicity
that the capacitance matrix $C_{ij}$ is diagonal one arrives at the
AES action $S_{\rm AES}[\phi ]$~\cite{Ambeg},
\begin{equation}
S_{\rm AES}[\phi
]=\frac{1}{4E_{c}}\sum\limits_{i}\int\limits_{0}^{\beta }d\tau \left(
\frac{d\phi _{i}}{d\tau }\right) ^{2}+\frac{g_{T}}{2}
\sum\limits_{i,j}\int\limits_{0}^{\beta }d\tau d\tau ^{\prime }\alpha
(\tau -\tau ^{\prime })\left( 1-\cos \left( \phi _{ij}(\tau )-\phi
_{ij}(\tau ^{\prime })\right) \right) , \label{S}
\end{equation}
where
\begin{equation}
\alpha (\tau -\tau ^{\prime })=\frac{T^{2}}{\sin ^{2}\pi T(\tau -\tau
^{\prime })} .  \label{s100}
\end{equation}
In Eq.~(\ref{S}) $E_{c}$ is the charging energy $E_{c}=e^{2}/2C_{ii}$.
The second term in Eq.~(\ref{S}) contains the sum over neighboring
grains and $t$ is the tunneling energy between them.  In the limit of
small phase fluctuations we can expand the second term in
Eq.~(\ref{S}) with respect to $\phi _{ij}$ and obtain
\begin{equation}
S[\phi ]=\frac{1}{4E_{c}}\sum\limits_{i}\int\limits_{0}^{\beta }d\tau \left( 
\frac{d\phi _{i}\left( \tau \right) }{d\tau }\right) ^{2}+
\frac{g_{T}}{2} \sum\limits_{i,j}\int\limits_{0}^{\beta }d\tau 
d\tau ^{\prime }\alpha (\tau -\tau ^{\prime })
\left( \phi _{ij}(\tau )-\phi _{ij}(\tau ^{\prime })\right)
^{2}.  \label{S2}
\end{equation}
Using the periodicity of all functions in $\beta $ we write the Fourier
expansion as 
\begin{equation}
\phi _{ij}(\tau )=T\sum\limits_{i\omega _{n}}\phi _{ij}(i\omega _{n})\exp
(-i\omega _{n}\tau ),\hspace{1cm}\alpha (\tau )=T\sum\limits_{i\omega
_{n}}\alpha (i\omega _{n})\exp (-i\omega _{n}\tau ),  \label{fur}
\end{equation}
where $\omega _{n}=2\pi nT$ are Matsubara frequencies. Hence we obtain for the
action 
\begin{equation}
S[\phi ]=\frac{1}{4E_{c}}\sum\limits_{i}T\sum\limits_{\omega
_{n}}\omega _{n}^{2}\mid \phi _{i}(\omega _{n})\mid
^{2}+\frac{g_{T}}{\pi } \sum\limits_{i,j}T\sum\limits_{i\omega
_{n}}\mid \omega _{n}\mid \mid \phi _{ij}(i\omega _{n})\mid ^{2}.
\label{S3}
\end{equation}
The second term in Eq.~(\ref{S3}) describing the tunneling is linear in
frequency and keeps its form down to $T=0$. Therefore, it was attributed to
quantum dissipation~\cite{Ambeg}. Performing a Fourier transformation in
space 
\begin{equation}
\phi _{i}(\omega _{n})=\sum\limits_{{\bf k}}\phi _{\omega _{n}}({\bf k})\exp
(i{\bf k}{\bf R}_{i}) ,
\end{equation}
we rewrite the action $S[\phi ]$ as 
\begin{equation}
S[\phi ]=\frac{1}{2}\sum\limits_{{\bf k}}T\sum\limits_{\omega
_{n}}\left( \frac{\omega _{n}^{2}}{2E_{c}}+\frac{4g_{T}}{\pi }|\omega
_{n}|\sum_{a}(1-\cos {\bf k}{\bf d}_{a})\right) |\phi _{\omega _{n}}
({\bf k})|^{2} \label{smallfluct} ,
\end{equation}
where ${\bf d}_{a}$, $a=x$, $y$, $z$ $\ $\ are vectors connecting the
centers of the neighboring grains. The phase-phase correlation function 
\[
\Pi _{\rm AES}\left( \omega _{n},{\bf k}\right) =T\langle \phi
_{\omega _{n}}( {\bf k})\phi _{\omega _{n}}^{\ast }({\bf k})\rangle
\]
can be immediately calculated using Eq.~(\ref{smallfluct}) and we obtain 
\begin{equation}
\Pi _{\rm AES}\left( \omega _{n},{\bf k}\right) =\frac{4E_{c}}
{\omega _{n}^{2}+\frac{8}{\pi }g_{T}E_{c}|\omega _{n}|
\sum_{a}(1-\cos {\bf k}{\bf d}_{a})} .
\label{cor}
\end{equation}

Suppose that one assumes, following Refs.\onlinecite{Zaikin,Fazio},
that Eq.~(\ref{S}) is applicable to a system of tunnel junctions down
to $T=0$.  This would imply in particular that Eq.~(\ref{cor}) is
correct for $T \to 0$, too.  However, the derivative $d\phi _{i}/d\tau
$ is proportional to the voltage on the grains which in turn is
linearly related to the charge density. Multiplying Eq.~(\ref{cor}) by
$\omega _{n}^{2}$, one thus obtains the density-density correlation
function $K\left( \omega _{n},{\bf k}\right) $.  Let us consider limit
of small frequencies and wave vectors $k$. In this limit, the network
should correspond to a bulk normal metal, where one should have the
propagator corresponding to a screened Coulomb interaction. However,
the limit $\omega $, $k\rightarrow 0$ in Eq. (\ref{cor}) does not
reproduce the screened Coulomb interaction.

How to resolve this discrepancy? In the next section we carry out an
expansion in $t_{ij}$ and demonstrate that keeping only the second
order, which led us to Eq.~(\ref{S}) is not sufficient and that, in
the limit of small frequencies or temperatures, one should sum up an
infinite class of diagrams. In section \ref{sigma} we will then show
how these processes can be included into an effective action
formulation structurally similar to the AES approach.

\section{Screened Coulomb interaction in granular metals}
\label{metal}

In this section we present a detailed derivation of the
density-density correlation function $K\left( \omega _{n},{\bf
k}\right) $ summing an infinite series of diagrams. For the
computation we use the standard diagram technique~\cite{AGD} with a
modification suggested in Refs.~\onlinecite {Bel,Belobor} to include
tunneling between the grains. Within this technique one should make an
expansion in the tunneling $t_{ij}$, the phases $\phi _{i} $ and in
the impurity potential $u_{i}\left( r\right) $. Below we denote the
electron Green functions by solid lines, the phases $\phi _{i}$ by
wavy lines and the tunneling elements $t_{ij}$ by crossed circles.
Impurity propagators corresponding to Eq. (\ref{a3}) are denoted by
dashed lines.

\begin{figure}
\epsfysize=2cm
\centerline{\epsfbox{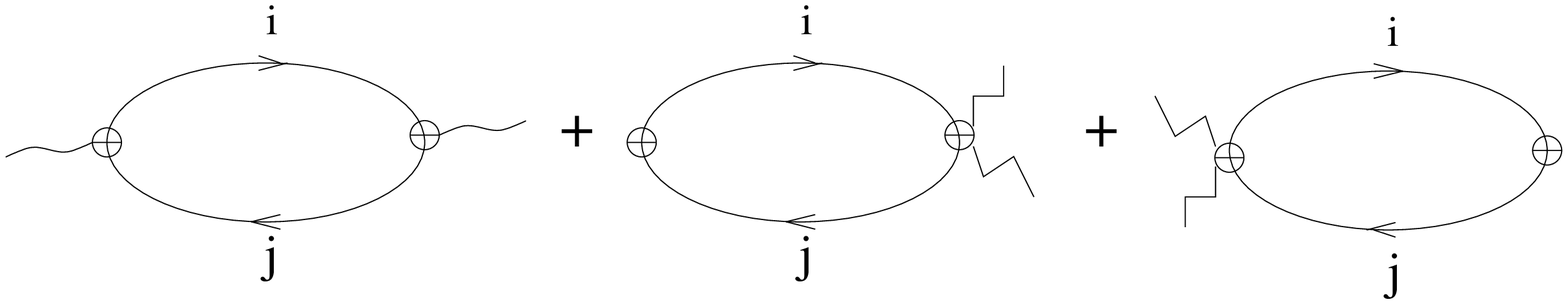}}
\caption{Polarization operator for a granular metal in the lowest
order in tunneling.}
\label{grainver}
\end{figure}

Before turning to the quantitative discussion, let us briefly outline
the skeleton of the analysis. To lowest order in the tunneling matrix
elements, the polarization operator of the system is represented by
the process depicted in Fig. \ref{grainver}. This diagram describes
the (absolute square of) amplitude for tunneling from one grain
into the next and the subsequent relaxation inside the 'target grain'
(represented through the Green function). The quantitative evaluation
of this process leads to the coupling constant $g_T$ of the AES
approach. 

\begin{figure}
\epsfysize=5cm
\centerline{\epsfbox{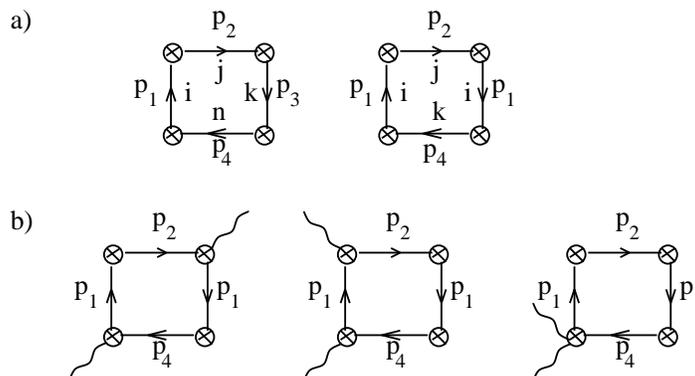}}
\caption{All classes of diagrams which contribute to the polarization
operator of granular metals in the fourth order of tunneling.}
\label{hightun}
\end{figure}

\begin{figure}
\epsfysize=7cm
\centerline{\epsfbox{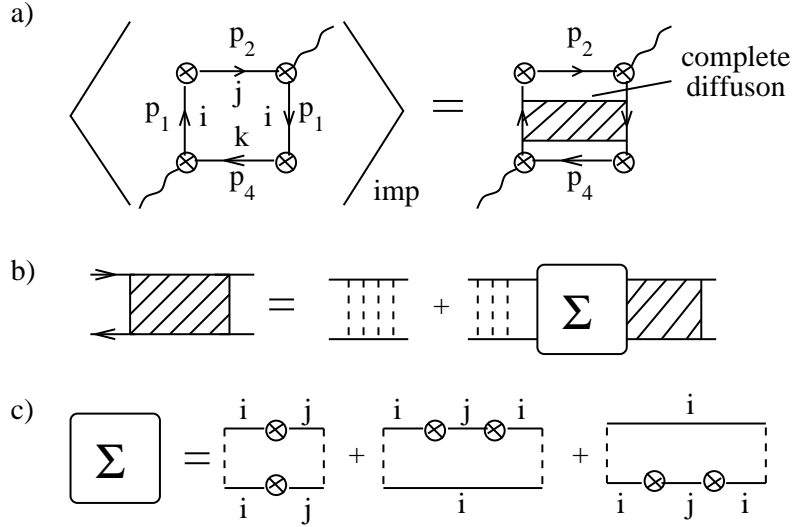}}
\caption{Equation for complete diffuson of granular metals.}
\label{diffuson}
\end{figure}

On this level, coherence effects associated to mulitiple tunneling
and/or impurity scattering are completely neglected; the tunneled
electron 'forgets' about its phase memory implying that the
dissipative picture of the AES approach obtains. Indeed, the two Green
functions depicted in Fig. \ref{grainver} live in different grains
which means that no phase correlation is possible. How can this
picture change in principle? Including one more order in the
tunneling, leads to corrections of the structure shown in
Fig. \ref{hightun}. What makes these processes qualitatively
different from those discussed above is that they include the
possibility of {\it phase coherent} multiple impurity scattering: The
two Green functions labeled $i$ or $j$ scatter off the same impurities
and/or chaotic potential where the associated scattering phases may
cancel due to the fact that one of the Green functions is advanced,
the other retarded. The net two-particle modes emanating from such
processes, 'diffusons' in the context of disordered systems, are
represented through the hatched regions of Fig. \ref{diffuson},
right. The key question now is, are such processes relevant or not? A
crude estimate can be given as follows: It is known that for low
temperatures (for a precise definition of 'low', see below), each
diffuson mode diverges as $\delta/T$. The fact that we are considering
processes of higher order in the tunneling amplitude introduces one
more power of $g_T$. Thus, for $T \sim \delta g_T$, higher order
processes become as important as the first order contribution.  For
lower temperatures, these processes {\it must} be taken into account
to obtain a physically correct picture. In the following we will put
this discussion onto a firm basis and discuss how the coherent
tunneling series can be summed in a controlled way.

\begin{figure}
\epsfysize=2cm
\centerline{\epsfbox{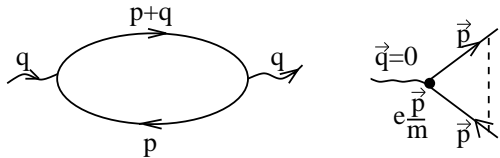}}
\caption{Diagram for the Drude conductivity and current vertex for a
bulk metal.}
\label{vertex}
\end{figure}

Before starting the computation for a granular system we want to
mention an important difference between the diagrams for the
current-current and density-density correlation functions. Calculating
the classical conductivity for a granular system in the lowest order
in the tunneling we need not renormalize the current vertices by
impurities. To understand this fact let us recall a well known result
for bulk metals. The diagram for the Drude conductivity is shown in
Fig. \ref{vertex}.
The second diagram in Fig.~\ref{vertex}, renormalizing the current vertex,
gives an additional integral over momentum that yields zero, 
\begin{equation}
\frac{e}{m}\int {\bf p}G_{0\varepsilon }({\bf p})G_{0\varepsilon }
({\bf p})d^{3}p=0  . \label{b1}
\end{equation}
In Eq.~(\ref{b1}), the Green function $G_{0\varepsilon }
\left( {\bf p}\right) $ is determined by Eq.~(\ref{a15}). 
Diagrams with many impurity
lines yield zero as well and the current vertices are not renormalized.
The same is true for granular metals, which can be seen after a proper
replacement of the current 
\begin{equation}
e\frac{{\bf p}}{m}\rightarrow et{\bf d}\sin {\bf p}{\bf d}  \label{b2} .
\end{equation}
As concerns the polarization operator or the density-density correlation
function, the situation is different. In this case, instead of vector
vertices we have scalar ones and the renormalization due to 
impurities can be important.

\begin{figure}
\epsfysize=1.2cm
\centerline{\epsfbox{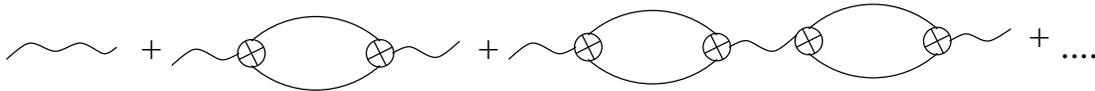}}
\caption{Diagrammatic analogy of AES action in the limit of small phase
fluctuations.}
\label{series}
\end{figure}

Expanding the functional integral, Eq.~(\ref{a11}) in the tunneling term 
$S_{1}$, Eq. (\ref{a13}), we obtain to second order in $t_{ij}$ and in 
$\phi _{i}$ the diagrams represented in Fig. \ref{grainver}. 
The two electron lines relate to different grains. As the impurities
in the different grains are not correlated, impurity lines connecting
two Green functions are do not exist (although the diagram represents
a polarization loop.) The second and third diagrams in
Fig.~\ref{grainver} are equal to each other and have an opposite sign
with respect to the first diagram. The analytical result for the sum
of the three diagrams reads
\begin{equation}
P_{0}\left( \omega_n ,{\bf k}\right) =-|\omega_n| (2/\pi) g_{T}
\sum\limits_{a=1}^3\left( 1-\cos k_ad\right) ,
\label{b3}
\end{equation}
where the dimensionless tunneling conductance $g_{T}$ is given by
Eq.~(\ref{Sg}).  We see from Eq.~(\ref{b3}) that, although the Green
functions $G_{0\varepsilon }\left( {\bf p}\right) $ determined by
Eq.~(\ref{a15}) contain $\tau _{\rm imp}$, all information about the
disorder drops out of the polarization loop.  To calculate the
phase-phase correlation function one should notice that the bare
propagator corresponding to $S_{2}$, Eq. (\ref{a14}), has the form
\begin{equation}
\Pi _{0}\left( \omega _{n},{\bf k}\right) =\frac{4E_{c}}{\omega _{n}^{2}}
\label{b4}
\end{equation}
and diverges at $\omega \rightarrow 0$. Therefore, one should sum up the
infinite geometrical series represented in Fig. \ref{series}. 
The summation can be easily performed and we arrive at the propagator
$\Pi _{\rm AES}\left( \omega ,{\bf k}\right) $, Eq.~(\ref{cor}).
Summarizing, we conclude that, in order to reproduce the results
obtained in the quadratic approximation of the AES action,
Eqs.~(\ref{S2}, \ref{S3}, \ref{smallfluct}), one should sum up the
diagrams of Fig.~\ref{series}.  On the other hand, the form of the
propagator $\Pi _{\rm AES}\left( \omega _{n},{\bf k}\right) $ does not
correspond to the propagator of a normal metal with a screened Coulomb
interaction (see {\em e.g.}~\onlinecite{Altshuler}), which signals
that something is missing in this analysis.

To resolve this paradox one should consider diagrams of higher order in
tunneling, which complicates the calculations. First of all, let us
note that in order to consider higher order diagrams one should specify
tunneling more explicitly. We assume that the radius of the area at which
the grains contact each other much exceeds atomic distances (large number of
conducting channels). At the same time, the potential barrier between the
grains can be arbitrarily large, which can lead to an arbitrary tunneling
conductance. Such a situation can be conveniently modeled by random
tunneling matrix elements $t_{pq}$ correlated as 
\begin{equation}
\langle t_{pq}t_{p^{\prime }q^{\prime }}\rangle =t^{2}(\delta
_{pp^{\prime }}\delta 
_{qq^{\prime }}+\delta _{pq^{\prime }}\delta _{p^{\prime }q}) ,
\label{average}
\end{equation}
where $t_{pq}$ are written for the same contact. Correlations of the
matrix elements of different contacts are assumed to be zero. To
simplify notations we do not write subscripts numerating the contacts.
Taking into account Eq.~(\ref{average}) is indeed very important. In
order to illustrate this fact we consider two diagrams represented in
Fig.~\ref {hightun}a. We see immediately the difference between them:
the first diagram is not averaged over $t_{pq}$ and can contains four
different momenta. After averaging over $t_{pq}$ one obtains the
second diagram containing only three different momenta.

As it has been mentioned, we will expand not only in the tunneling amplitude
$t $ but also in the phases $\phi $. Performing 
the expansion in the phases and
denoting them by wavy lines we have three different possibilities as depicted
in Fig. \ref{hightun}b. 
Now we can start averaging over impurities. As usual~\cite{AGD}, we
average first the Green functions and obtain Eq.~(\ref{a15}) for the average
Green function $G_{0\varepsilon }\left( {\bf p}\right) $. Then, we can
consider averages with impurity lines connecting the Green functions on
opposite sides of the diagrams in Fig.~\ref{hightun}b. Diagrams with
non-intersecting impurity lines are most important and we recognize the 
well-known diffusons.

In the present paper we consider the limit of sufficiently small grains,
such that all relevant energies like temperature $T$, frequency $\omega $,
tunneling energy $t$, etc. are much smaller than the Thouless energy 
$E_{T}=\pi ^{2}D_{0}/R^{2}$,where $D_{0}$ is the diffusion coefficient of a
single grain and $R$ is the radius of the grain. In this limit the spectrum
of the diffuson in a single grain is zero-dimensional (no dependence on
momenta inside the grain). Coupling between the grains leads to a dependence
on quasi-momentum ${\bf k}$. Diagrams that should be summed in order to give
the complete form of the diffuson are represented in Fig.~\ref{diffuson}a.

Now let us present some details of the calculation of the diagrams in 
Fig.~\ref{hightun}. The different classes of diagrams contributing to the
polarization operator of the granular metals before impurity averaging
are shown in Fig.~\ref{hightun}b. Before impurity averaging the sum
over Matsubara frequencies for the first diagram in Fig.~\ref{hightun}b is 
\begin{equation}
T\sum\limits_{\varepsilon _{n}}G(i\varepsilon _{n}+i\omega _{n},{\bf p}_1)
G(i\varepsilon _{n}+i\omega _{n},{\bf p_{2}})G(i\varepsilon _{n},{\bf p}_1)
G(i\varepsilon _{n},{\bf p}_4) . \label{GF}
\end{equation}
Writing this expression we used Eq. (\ref{average}).  After impurity
averaging of the first diagram in Fig.~\ref{hightun}b, we obtain the
diagram represented in Fig.~\ref{diffuson}a.

Now the disorder averaging can be done and we obtain, in a standard
way, the average of the product of two Green functions for a granular
metal, $\langle G^{R}G^{A}\rangle =D(\omega_n ,{\bf q})$. Here
\begin{equation}
D=D^{(0)}+D^{(0)}\Sigma D,  \label{Dyson}
\end{equation} 
$D^{(0)}(\omega_n, {\bf k})=2\pi \nu /|\omega_n|$ being the diffuson
for a single isolated grain.  The equation for the complete
diffuson~\cite{Belobor} is shown in Fig.~\ref{diffuson}b.  The
self-energy $\Sigma $ in Eq.~(\ref{Dyson}) is the sum of the three
diagrams in Fig.~\ref{diffuson}c. The second and third diagrams in
Fig.~\ref {diffuson}c are equal to each other and have the opposite
sign with respect to the first diagram. Solving Eq.~(\ref{Dyson}) we
obtain
\begin{equation}
D(\omega_n,{\bf k})=\frac{2\pi \nu }{|\omega_n| +(2/\pi) g_{T}\delta 
\sum\limits_{a=1}^{3}(1-\cos k_{a}d)} .  \label{Dyson1}
\end{equation}

\begin{figure}
\epsfysize=2.5cm
\centerline{\epsfbox{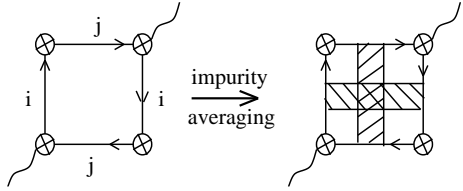}}
\caption{Example of a high order correction to the polarization of granular
metals.}
\label{dif}
\end{figure}

The calculation of the second diagram in Fig.~\ref{hightun}b is
analogous to the calculation of the first one. The third diagram in
Fig.~\ref{hightun}b is independent ofthe frequency $\omega _{n}$ and
yields a constant.  The diagram in Fig.~\ref{diffuson}a contains only
one diffuson. More complicated diagrams can be drawn and one of them
is depicted in Fig.~\ref{dif}.
However, all such many-diffuson contributions can be neglected provided the
tunneling conductance $g_{T}$, Eq.~(\ref{Sg}), is large . A direct
calculation of this diagram shows that it involves the small parameter 
$1/g_{T}$. This is similar to expansions in the diffusion modes for a bulk
metal where the inverse conductivity is the expansion parameter. The
condition $g_{T}\gg 1$ means that we are far from the Anderson
metal-insulator transition.

The result of averaging and summation of the diagrams represented in 
Fig.~\ref{hightun} can be written as 
\begin{equation}
P_{1}(\omega _{n},{\bf k})=\frac{|\omega_n|\delta\left ((2/\pi
)g_{T}\sum\limits_{a}(1-\cos k_{a}d)\right)^2}{|\omega _{n}|+(2/\pi
)g_{T}\delta \sum\limits_{a}(1-\cos k_{a}d)} .
\label{newdif}
\end{equation}
We note here that the functions $P_0$ and $P_1$ have a different
sign. Comparing the function $P_{1}\left( \omega_n,{\bf k}\right) $,
Eq. (\ref {newdif}), with the quantum dissipation part
$P_{0}\left( \omega_n,{\bf k}\right) $, Eq. (\ref{b3}), represented in
Fig.~\ref{grainver} we see immediately that the contribution
$P_{1}\left( \omega_n,{\bf k}\right) $, Eq. (\ref{newdif}), can be
neglected only if the temperature $T$ is sufficiently high $T\sim
\left| \omega _{n}\right| \gg g_{T}\delta $ (we recall that the static
component of the phase $\phi $ has been neglected and therefore
$\omega _{n}\neq 0$ in Eq.~(\ref{newdif}). So, the notion of quantum
dissipation is applicable essentially at non-zero temperatures only
for a granular material. However, it can be used for a contact connecting 
two bulk metals with an infinite volume.

\begin{figure}
\epsfysize=2.5cm
\centerline{\epsfbox{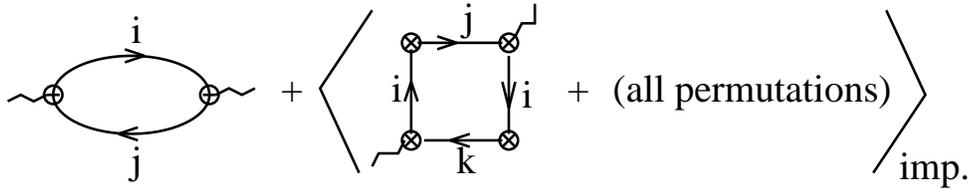}}
\caption{Relevant polarization loops for a granular metal.}
\label{complete}
\end{figure}

At low temperatures $T\leq g_{T}\delta $, there is no reason to neglect the
contribution $P_{1}\left( \omega _{n},{\bf k}\right) $, Eq. (\ref{newdif}),
and its presence changes completely the form of the propagator. Adding the
functions $P_{0}\left( \omega _{n},{\bf k}\right) $, Eq. (\ref{b3}), and 
$P_{1}\left( \omega_n ,{\bf k}\right) $, Eq. (\ref{newdif}), we obtain the
total polarization $P\left( \omega_n ,{\bf k}\right) $
\begin{equation}
P\left( \omega_n ,{\bf k}\right) =-\frac{|\omega_n|^2(2/\pi)g_T
\sum \limits_{a}(1-\cos k_ad)}{|\omega_n|+(2/\pi)g_T\delta
\sum\limits_{a}(1-\cos k_ad)} .
\label{total}
\end{equation}
Proper diagrams contributing to this function are represented in Fig.~\ref
{complete}.
Now, we can write the final expression for the phase correlation function 
$\Pi \left( \omega _{n},{\bf k}\right) $ that should be written instead of 
$\Pi _{\rm AES}\left( \omega _{n},{\bf k}\right) $, Eq. (\ref{cor}). Using the
bare propagator $\Pi _{0}\left( \omega _{n},{\bf k}\right) $, Eq.~(\ref{b4}), 
and the self-energy part $P\left( \omega _{n},{\bf k}\right) $, Eq.~(\ref
{total}), we write the phase propagator 
$\Pi \left( \omega _{n},{\bf k}\right)$ in the form 
\begin{equation}
\Pi \left( \omega _{n},{\bf k}\right) =\frac{\Pi_0}{1-\Pi_0
P}=\frac{4E_c/\omega_n^2}{1+4E_c\frac{(2/\pi)g_T
\sum\limits_a(1-\cos
k_ad)}{|\omega_n|+(2/\pi)g_T\delta\sum\limits_a(1-\cos k_ad)}}  .
\label{f1}
\end{equation}

Using the relation between the phase $\phi $ and the voltage $V$, Eq.~(\ref
{a10}), we derive an equation for the effective Coulomb interaction 
\begin{equation}
V_{\rm eff}\left( \omega _{n},{\bf k}\right) =\omega _{n}^{2}\Pi \left(
\omega _{n},{\bf k}\right) .  \label{f2}
\end{equation}
On the other hand, the dynamically screened Coulomb interaction in a
disordered bulk metal can be written in the form 
\begin{equation}
\tilde{V}_{\rm eff}(\omega_n, {\bf k})=\frac{\tilde{V}_0(\bf
k)}{1+\tilde{V}_0({\bf k})\frac{\nu D{\bf k}^2}{|\omega_n|+D{\bf
k}^2}} ,\label{f3}
\end{equation}
where $\tilde{V}_0$ is the bare Coulomb potential. We see that
Eqs.~(\ref{f1}, \ref{f2}) written in the continuum limit correspond to
Eq.~(\ref{f3}).

Thus, we conclude that, when calculating physical quantities, the main
contribution is due to small $k$ only in the limit $T\ll g_{T}\delta $.
Provided this inequality is satisfied, the properties of the tunnel 
junction array are
equivalent to the properties of a weakly disordered bulk metal with Coulomb
interaction. In the opposite limit $T\gg g_{T}\delta $, one can describe the
system in the ``quantum dissipation'' approximation.

Calculations were performed in this section using a diagrammatic expansion
in the lowest order in the field $\phi $. On the other hand, the AES free
energy functional, Eq.~(\ref{S}), is written for arbitrary values of 
$\phi $. Can one generalize Eq.~(\ref{f2}) to arbitrary values of $\phi $? 
We will
try to do this in the next section, deriving a simplified version of a
replica $\sigma $-model and using a non-trivial saddle point.

\bigskip

\section{Effective action for granular metals with Coulomb interaction.}

\label{sigma}

In the preceding sections we analyzed effects of the Coulomb interaction on
the electron motion in granular materials using a diagrammatic expansion.
This approach works well when fluctuations of the phase $\phi $ or, in other
words, fluctuations of the voltages on the grains are small. However, 
one can imagine situations when the fluctuations are large. For
example, one can try to consider non-perturbative excitations like
instantons~\cite{korshunov} corresponding to an integer charge transfer
between grains.
In principle, the AES action, Eq. (\ref{S}), can describe 
such excitations
very well but, as we have seen previously, it can only be used at
sufficiently high temperatures $T\gg g_{T}\delta $. So, our task now is 
to derive an action that will be applicable at lower temperatures.

The derivation at temperatures $T\leq g_{T}\delta $ is more
complicated because now we should explicitly take into account
disorder, which was not necessary for the derivation of the AES action
(strictly speaking, disorder determines the mean free time $\tau _{\rm
  imp}$ in the Green functions used for derivation of the AES
functional but this is a trivial contribution).  On the other hand,
the problem is not as complicated as the problem of Anderson
localization in the presence of interactions considered in
Refs.~\onlinecite{Fin,Fin82}.  In our diagrammatic expansions we
considered the limit of large conductances $g_{T}\gg 1$ and neglected
all weak localization corrections.

Nevertheless, we start the derivation of the action using the replica 
$\sigma $-model approach for interacting systems suggested by Finkelstein 
\cite{Fin,Fin82}. Necessary simplifications will be made in the 
$\sigma $-model. Although there are difficulties in using the replica 
$\sigma $-models for non-perturbative calculations, our goal is more 
modest and
the final results obtained in the limit $g_{T}\gg 1$ neglecting all weak
localization corrections will not depend on replica indices at all.

The model under consideration has been formulated in Section~\ref{choice}.
The derivation of the $\sigma $-model can be carried out starting from 
Eqs.~(\ref{a1}-\ref{a5}). Instead of the fields $\psi $ and $V$ one should 
write fields $\psi =\{\psi ^{a}\}$ and $\{V^{a}\}$ carrying the replica index 
$a=1,....r.$ Upon calculating physical quantities for an arbitrary $r$, 
one should put $r=0$.

A proper $\sigma $-model for bulk systems with interaction has been derived
in Refs.~\onlinecite{Fin,Fin82}. A generalization to granular metals can been
made without difficulties following Ref.~\onlinecite{Efetov}, where a 
$\sigma $-model was written for a granular metal without interaction. 
Although one
can proceed along the lines of Refs.~\onlinecite{Fin,Fin82}, {\em i.e.}
expanding the
action obtained after integration over $\psi $ in the field $V$, a more
economic way is to use the substitution given by 
Eqs.~(\ref{a90}, \ref{a10}), as well as the
subsequent Eqs.~(\ref{a11}-\ref{a14}). Of course, this is already an
approximation because we neglect the static component of the phase $\phi $.
However, for our purposes this is not an essential restriction because we do
not want to consider effects of localization. Moreover, working in the limit 
$T\gg \delta $ or $g_{T}\gg 1$ allows us to ignore the static 
component of $\phi $.

The derivation of the $\sigma $-model starting from Eqs.~(\ref{a11}-\ref{a14})
is practically the same as the one presented in Ref.~\onlinecite{Efetov}. As
usual, one decouples the ``effective interaction'' of the type $\psi ^{4}$
that appears after averaging over impurities by a Gaussian integration over
matrices $Q$. These matrices contain as elements both the replica
indices and Matsubara frequencies (or two imaginary times $\tau $ and $\tau
^{\prime }$). Neglecting the tunneling term, Eq.~(\ref{a13}), one would
obtain a zero-dimensional $\sigma $-model (we assume that all relevant
energies are smaller than the Thouless energy $E_{T}$ of a single grain). The
relevant tunneling term in the $\sigma $-model is obtained by a cumulant
expansion in the tunneling. The only difference with respect to the model
without interaction of Ref.~\onlinecite{Efetov} is the presence of the phases 
$\phi $ in Eq.~(\ref{a13}), which leads to additional phase factors in the
term describing the coupling between the grains.

The final result for the free energy $F[Q,\phi ]$ can be written as
\begin{eqnarray}
&&\hspace{0.7cm}F[Q,\phi ]=F_{2}[\phi ]+F_{{\rm \omega }}+F_{{\rm
T}}[Q,\phi ], \label{full5} \\ &&\qquad F_{2}[\phi
]=\frac{1}{2e^2}\sum_{i,j}\int {\rm tr}C_{ij}\frac{\partial \phi
_{i}\left( \tau \right) }{\partial \tau }\frac{\partial \phi
_{j}\left( \tau \right) }{\partial \tau }d\tau , \label{full6} \\
&&\qquad F_{\omega }[Q]=\frac{\pi }{\delta }\sum_{i}\int {\rm tr}
(\hat{\omega}_{\tau }Q_{i}\left( \tau ,\tau \right) )d\tau ,
\label{full7} \\ &&\qquad F_{{\rm T}}[Q,\phi
]=-\frac{g_{T}}{4}\sum_{i,j} \int {\rm tr}\left( e^{i\phi _{ij}(\tau
)}Q_{i}\left( \tau ,\tau ^{\prime } \right) e^{-i\phi _{ij}(\tau
^{\prime })}Q_{j}\left( \tau ^{\prime },\tau \right) \right) d\tau
d\tau ^{\prime }, \label{full8}
\end{eqnarray}
where $\phi _{ij}\left( \tau \right) =\phi _{i}\left( \tau \right) -\phi
_{j}\left( \tau \right) $ and the symbol {\rm tr }implies a trace over
both
replica indices. The field $\phi $ is diagonal in the replica indices, while
the matrix $Q\left( \tau ,\tau ^{\prime }\right) $ is a $2r\times 2r$ matrix
with the constraints 
\begin{equation}
\int Q\left( \tau ,\tau ^{\prime \prime }\right) Q\left( \tau ^{\prime
\prime },\tau ^{\prime }\right) d\tau ^{\prime \prime }=\delta \left( \tau
-\tau ^{\prime }\right) ,\text{ \ \ {\rm tr}}Q\left( \tau ,\tau \right) =0 .
\label{e1}
\end{equation}

The action of the operator $\hat{\omega}$ in Eq.~(\ref{full7}) on an arbitrary
function $f\left( \tau ,\tau \right) $ is given by the relation 
\begin{equation}
\hat{\omega}f\left( \tau ,\tau ^{\prime }\right) =-i\lim_{\tau
^{\prime }\rightarrow \tau }\frac{\partial }{\partial \tau }f\left( \tau
,\tau ^{\prime }\right)  .\label{e2}
\end{equation}
In frequency representation this operator is equal to a vector of 
fermionic Matsubara frequencies.
The matrix $Q\left( \tau ,\tau ^{\prime }\right) $, satisfying the
constraints, Eq. (\ref{e1}), can be conveniently parametrized as 
\begin{equation}
Q\left( \tau ,\tau ^{\prime }\right) =\int U\left( \tau ,\tau ^{\prime
\prime }\right) \Lambda _{\tau ^{\prime \prime },\tau ^{\prime \prime \prime
}}\bar{U}\left( \tau ^{\prime \prime \prime },\tau ^{\prime }\right) d\tau
^{\prime \prime }d\tau ^{\prime \prime \prime }  , \label{e3}
\end{equation}
where 
\begin{equation}
\Lambda _{\tau ,\tau ^{\prime }}=\frac{iT}{\sin T\pi \left( \tau -\tau
^{\prime }\right) }  \label{e4}
\end{equation}
and $U,\bar{U}$ are unitary matrices 
\begin{equation}
\int U\left( \tau ,\tau ^{\prime \prime }\right) \bar{U}\left( \tau ^{\prime
\prime },\tau ^{\prime }\right) d\tau ^{\prime \prime }=\delta \left( \tau
-\tau ^{\prime }\right).  \label{e5}
\end{equation}
The matrix $\Lambda $ in frequency representation has the form 
\[
\Lambda _{mn}=\delta _{mn} {\rm sgn}\left( \varepsilon _{n}\right)  .
\]
The limiting cases of the free energy $F[Q,\phi ]$, Eqs.~(\ref{full5}-\ref
{e2}) are rather simple. Without electron-electron interaction the phase
difference between neighboring grains equals to zero $\phi _{ij}=0$.
Such an action for granular metals corresponds to that derived within the
supersymmetry scheme by one of the authors~\cite{Efetov}. The second
limiting case is achieved at sufficiently high temperatures. In this
limit disorder is not important and the matrix $Q$ does not fluctuate, being
equal to the value $\Lambda $, Eq.~(\ref{e4}). Inserting $Q=\Lambda $ into
Eqs.~(\ref{full5}-\ref{full8}) we reproduce immediately the AES action, 
Eq. (\ref{S}) (in fact we obtain the $r$ times replicated AES functional but
only one replica field is important for us).

At not too high a temperature, Eqs.~(\ref{e3}-\ref{e5}) lead to a
non-trivial behavior even if the disorder is weak and all effects
related to weak localization can be neglected. The limit of weak
disorder enables us to neglect fluctuations of $Q$ at given $\phi $
and take its value from a saddle-point equation.  We thus reduce the
computation of the functional integral with the free energy functional
to the solution of a saddle-point equation, substituting the solution
to the free energy $F[Q,\phi ]$, Eqs.~(\ref{full5}-\ref{full8}),
calculating a functional integral with the reduced free energy
$\bar{F}[\phi ]$
\begin{equation}
\bar{F}[\phi ]=F[\bar{Q},\phi ]  \label{e6}
\end{equation}
where $\bar{Q}$ is the solution of the saddle-point equation for a given 
$\phi $.

Minimizing the functional $F[Q,\phi ]$, Eqs. (\ref{full5}-\ref{full8}), with
the constraints, Eqs.~(\ref{e1}), we obtain the following equation 
\begin{equation}
\frac{g_{T}}{2}\sum_{j}\int [e^{i\phi _{ij}\left( \tau \right)
}\bar{Q} _{i}\left( \tau ,\tau ^{\prime \prime }\right) e^{-\phi
_{ij}\left( \tau ^{\prime \prime }\right) },\bar{Q}_{j}\left( \tau
^{\prime \prime },\tau ^{\prime }\right) ]d\tau ^{\prime \prime
}+{\frac{\pi }{\delta }} [\bar{Q}_{i}\left( \tau ,\tau \right)
,\hat{\omega})]=0, \label{e7}
\end{equation}
where the summation over $j$ is performed over the nearest neighbors of $i$.
In the commutator $[..,..]$ one should exchange properly the times when
changing the order of the functions. 

The Eq. (\ref{e7}) represents an integral equation that enables us to
find, in principle, the solution $K_{i}\left( \tau \right) $ for
arbitrary values of the parameter $g_{T}\delta /T$. In general, this
equation can be solved only numerically. However, for $g_T\gg 1$
drastic simplifications arise.  For large intergranular coupling, the
phase $\phi$ fluctuates only weakly from grain to grain which means
that $\phi_{ij}$ is small. Under these conditions, a solution of the
diagonal form
\begin{equation}
\bar{Q}_{i}=e^{iK_{i}\left( \tau \right) }\Lambda _{\tau ,\tau ^{\prime
}}e^{-iK_{i}\left( \tau ^{\prime }\right) }  \label{e8}
\end{equation}
can be sought for. As the solution is assumed to have a diagonal form,
all replica indices decouple and the equations can be solved for each
replica separately.  Therefore we can drop the replica indices and
consider Eq.~(\ref{e7}) as an equation for a single function
$K_{i}\left( \tau \right) $ for a given function $\phi _{ij}\left(
  \tau \right) $.

In the high temperature limit $T\gg g_{T}\delta $  the second term in 
Eq.~(\ref{e7}) is much larger than the first one and we arrive at the solution 
\begin{equation}
K_{i}\left( \tau \right) =0 , \label{e9}
\end{equation}
which leads to the AES free energy functional, Eq.~(\ref{S}).

In the opposite limit of low temperatures $T\ll g_{T}\delta $ Eq.~(\ref{e7})
can be simplified using the assumption that $K_{i}\left( \tau \right) $
varies slowly in space. Then, one may expand the exponentials in 
$K_{ij}\left(
\tau \right) =K_{i}\left( \tau \right) -K_{j}\left( \tau \right) $, which
leads to the equation 
\begin{equation}
g_{T}\sum_{j}\Lambda \lbrack \left( K_{ij}+\phi _{ij}\right) ,\Lambda ]
+
\frac{i\pi }{\delta }[\frac{\partial }{\partial \tau },K_{i}]=0 .  \label{e10}
\end{equation}

Eq.~(\ref{e10}) is presented without writing the integration over
imaginary times $\tau $ explicitly, but of course this integration is
implied.  Transforming this equation to frequency and momentum space
we obtain
\begin{eqnarray}
\left(2g_{T}\sum\limits_{l=1}^{3}(1-\cos q_{l}d)(1-{\rm
sgn\,}(\omega_n){\rm sgn\,}(\omega_m))+{\frac{\pi }{\delta }}(\omega
_{n}-\omega _{m})({\rm sgn\,}(\omega_n)- {\rm sgn\,}(\omega_m)\right)
K_{q,n-m} =
\label{e11} \\ -{\frac{i\pi }{\delta }}V_{q,n-m}({\rm sgn\,}(\omega_n)-{\rm
sgn\,}(\omega_m)), \nonumber
\end{eqnarray}
which gives the solution
\begin{equation}
K_{k,n}=-V_{k,n}i{\frac{{\,{\rm sgn\,}}\omega_n}{(2/\pi )g_{T}\delta
\sum\limits_{a=1}^{3}(1-\cos kd_a)+|\omega _{n}|}},  \label{e12}
\end{equation}
where $K_{k,n}$ and $V_{k,n}=(i/e)\omega_n \phi_nk$ are the Fourier
transforms of the functions $K_{i}\left( \tau \right) $ and
$V_{i}\left( \tau \right) $ respectively.  
Substituting this ansatz for $K$ back into the action we obtain 
\begin{equation}
F[V] = \frac{1}{4E_c} \sum_n \sum_k V_{k,n}\left(1+ 4E_c
\frac{(2/\pi)g_T\sum_a(1-\cos k_ad)}{|\omega_n|+(2/\pi)g_T\delta
\sum_a(1-\cos k_ad)}\right)V_{-k,-n}
\end{equation}
where $E_c$ is the charging energy. In the limit of small
characteristic momentum, $k\ll d^{-1}$, we recover the diffusively
screened Coulomb interaction in a disordered evironment.

From here one can proceed to the calculation of physical observables
such as the conductance, the density of states or others. To this end
one should introduce configurations $Q\equiv T^{-1} \bar Q T$
fluctuating around the mean field configuration discussed above. Next
one would follow the standard algorithm of doing calculations
within the $\sigma$-model approach: (i) express the quantity of
interest in terms of $Q$, (ii) compute the functional average over the
effective action $F[Q]=F[V,T]$ as good as is possible.

Let us outline how the connection between the enlarged AES-type
formulation, (\ref{full5}) - (\ref{full8}) and the Finkelstein approach
for interacting disordered media can be made explicit. To do so, we
subject our $Q$'s to a gauge transformation,
\begin{equation}
Q_i \to e^{i\phi_i} Q_i e^{-i\phi_i}.
\end{equation}
as a result (i) the hopping part of the action, $F_{\rm T}$ becomes
$\phi$-independent. However (ii) in the frequency part, $\hat
\omega_{\rm T} \to \hat \omega_{\rm T} + \partial_\tau \phi_i$.
Gaussian integration over $\partial_\phi$ then produces an effective
action $F[Q]$ which, after taking a continuum limit, is identified as
the action of the Finkelstein approach.

\section{Conclusion}

In this paper we constructed a theoretical framework to describe
electronic transport in arrays of tunnel junctions or granular metals,
in the limit of large inter-granule conductance, $g_T$. Both disorder
and the electron-electron interaction were taken into account. The
prime objective of this enterprise was to unify two large, and
seemingly non-overlapping theories of interacting metallic compounds:
the AES approach, focusing on charging phenomena in individual grains
and their impact on large scale transport behaviour, and the
Finkelstein theory of disordered interacting metals with its emphasis
on the interplay interaction/disorder. The key to the reconciliation
of these two approaches was to observe that for low enough
temperatures, $T<g_T\delta$, the effective action underlying the AES
theory becomes incomplete. The physical reason is that for low
temperatures the electrons and holes participating in the tunneling
processes between individual grains maintain their quantum phase
memory for a long time, largely in excess of the average tunneling
time. This means that the particle tunneling is not only accompanied
by charging and dissipation (as in the AES approach), but also by
other, more long ranged physical processes. Specifically, we
identified the quantum dissipation contained in the AES action as the
high temperature limit of the long ranged screened electron-electron
interaction, an observation first made in\cite{Alt}.  We re-emphasize
that these phenomena were not bound to the presence of a significant
disorder concentration; indeed, none of the results discussed above,
displayed dependence on some 'disorder concentration'. The only thing
that mattered was chaoticity of the electron motion on scales set by
the phase coherence length, a condition that is practically always met
in real life systems.

Technically, two different routes for including these processes into
the theory were proposed. One consisted of the perturbative summation
of diagram classes associated to multiple chaotic scattering. The
other, based on an effective action formulation, indentified what is
'missing' in the AES action.  Remarkably, this second approach readily
led to a unification of the AES approach and the Finkelstein nonlinear
$\sigma$-model for weakly disorderd interacting metals. With the
benefit of hindsight, this fact is easy to understand: It is a well
known fact in mesoscopic physics, that the long processes resulting
from multiple scattering can be interpreted as a certain type of
Goldstone modes. Within an effective action approach, these modes must
be described by an own degree of freedom, the $Q$-matrices of the
nonlinear $\sigma$-model.  Thus it is no surprise that we obtained an
action of the type (\ref{full5} - \ref{full8}), involving $Q$-fields
and the Coulomb phase fields of the AES approach, as an effective
description of the low energy phase of the system. For large
temperatures, the fluctuations of the $Q$-matrices became inessential
and the AES action was retrieved. In contrast, the continuum limit of
this action was identified as the Finkelstein model and the
$Q$-matrices smoothly integrated into the AES approach, without
leading to conceptual complications.

Let us make some remarks on potential experimental ramifications of
our findings.  It has been predicted in
Refs.~\onlinecite{Fazio,Mooij90} that two-dimensional normal arrays of
tunnel junctions should undergo a Kosterlitz-Thouless-Berezinskii
(KTB) phase transition at a temperature $T_c$ of the order of the
charging energy $E_c$. At low temperatures, $T < T_c$, the array
should be in an insulating state (Coulomb blockade), whereas for
$T>T_c$ the array is conducting. In particular, at $T \agt T_c$, the
conductivity $\sigma$ of the array is predicted to increase with
temperature according to a square-root cusp dependence, $\sigma (T)
\sim \exp (- 2b/\sqrt{T/T_c -1})$, where $b$ is a constant of order
unity.  The experiments~\cite{Mooij90,Tighe93,Yamada93,Kobayashi95}
mentioned in the Introduction searched for this KTB transition.  In
view of our results, the applicability of the KTB scenario should be
governed by the parameter $g_T \delta/T_c$.  The experiments of
Refs.\onlinecite{Mooij90,Tighe93,Kobayashi95} were done on arrays with
relatively large grains, such that $g_T \delta/T_c \ll 1$. Indeed, the
results obtained in Refs.~\onlinecite{Mooij90,Kobayashi95} were in
agreement with the KTB scenario. However, Ref.~\onlinecite{Tighe93}
found thermally activated behavior of $\sigma(T)$, rather than the
predicted square root cusp dependence on temperature.  On the other
hand, the experiment by Yamada {\em et al.}~\cite{Yamada93} was done
on arrays consisting of relatively small Cu grains with a size of
about $40 \AA$, such that $g_T\delta/T_c\sim 20$. Clearly this is
beyond the range of applicability of Refs. \onlinecite{Fazio,Mooij90}
and the KTB scenario should be treated with care in this
case. Remarkably, Ref.~\onlinecite{Yamada93} concludes good agreement
of the results with the theory~\cite{Fazio,Mooij90}.  Given these
discrepancies, we conclude that a careful analysis of the
temperature-dependent conductivity of junction arrays in the framework
of the theory presented in this paper would be of interest.

\section{ACKNOWLEDGMENTS}

The authors thank A. Andreev, R. Fazio and A. Tschersich for
helpful discussions. A support of the Graduiertenkolleg 384 and the
Sonderforschungsbereich 237 is greatly appreciated.

\end{document}